\documentclass[aps,prl,showpacs,floatfix,superscriptaddress,twocolumn]{revtex4}
\usepackage{graphicx}
\usepackage{amsmath,amssymb,amsthm}
\usepackage{times}
\usepackage{bm}
\usepackage[dvips]{color}
\usepackage[usenames,dvipsnames,svgnames,table]{xcolor}

\newcommand{\be}{\begin{eqnarray}}
\newcommand{\ee}{\end{eqnarray}}

\def\beq{\begin{equation}}
\def\eeq{\end{equation}}

\begin{document}

\title{Long-time behavior of the  momentum distribution during the sudden expansion of \\
a spin-imbalanced Fermi gas in one dimension}

\author{C.\,J. Bolech}
\affiliation{Department of Physics, University of Cincinnati, Cincinnati, OH 45221, USA}
\affiliation{Kavli Institute for Theoretical Physics, Kohn Hall, University of California, 
Santa Barbara, CA 93106, USA}
\author{F. Heidrich-Meisner}
\affiliation{Kavli Institute for Theoretical Physics, Kohn Hall, University of California, 
Santa Barbara, CA 93106, USA}
\affiliation{Department of Physics and Arnold Sommerfeld Center for Theoretical Physics,
Ludwig-Maximilians-Universit\"at M\"unchen, D-80333 M\"unchen, Germany}
\author{S. Langer}
\affiliation{Department of Physics and Arnold Sommerfeld Center for Theoretical Physics, 
Ludwig-Maximilians-Universit\"at M\"unchen, D-80333 M\"unchen, Germany}
\author{I.\,P. McCulloch}
\affiliation{Centre for Engineered Quantum Systems, School of Mathematics and Physics, 
The University of Queensland, St Lucia, Brisbane 4072, Australia}
\author{G. Orso}
\affiliation{Laboratoire Mat\'eriaux et Ph\'enom\`enes Quantiques, 
Universit\'e Paris Diderot-Paris 7 and CNRS, UMR 7162, 75205 Paris Cedex 13, France}
\author{M. Rigol}
\affiliation{Department of Physics, Georgetown University, Washington, DC 20057, USA}
\affiliation{Physics Department, The Pennsylvania State University, 104 Davey Laboratory,
University Park, Pennsylvania 16802, USA}

\begin{abstract}
We study the sudden expansion of spin-imbalanced ultracold lattice fermions with attractive interactions
in one dimension after turning off the longitudinal confining potential. We show that the momentum distribution
functions of majority and minority fermions quickly approach stationary values due to a 
 quantum distillation mechanism that results in a spatial separation of pairs and majority fermions. As a consequence,
Fulde-Ferrell-Larkin-Ovchinnikov (FFLO) correlations are lost during the expansion. Furthermore, we argue that the shape of the
stationary momentum distribution functions can be understood by relating them to the integrals of motion in
this integrable quantum system. We discuss  our results in the context  of proposals to
observe FFLO correlations, related to recent experiments
by Liao {\it et al.}, Nature {\bf 467}, 567 (2010).
\end{abstract}


\maketitle
 
The combination of strong correlations and quantum fluctuations makes one-dimensional (1D) systems 
the host of exotic phases and physical phenomena \cite{giamarchi,cazalilla_citro_review_11}. Those phases and phenomena,
in many occasions first predicted theoretically, have been observed in condensed matter experiments and have 
begun to be studied with ultracold atomic gases \cite{cazalilla_citro_review_11}. A  system of 
particular interest in recent years has been the spin imbalanced 1D Fermi gas. Following theoretical predictions 
\cite{orso07,hu07,feiguin07,guan07,casula08,kakashvili09,hm10}, its grand canonical phase diagram has  recently been 
investigated experimentally \cite{liao_rittner_10}. The major interest in this model comes from the fact that its entire 
partially polarized phase has been theoretically shown \cite{yang01,feiguin07,tezuka08,batrouni08,rizzi08,luescher08,casula08} 
(for a review, see \cite{feiguin11}) to be the 1D-analogue of the Fulde-Ferrell-Larkin-Ovchinnikov (FFLO) state 
\cite{fulde_ferrel_64,larkin_ovchinnikov_65}. The FFLO phase was introduced to describe 
a possible equilibrium state in which magnetism and superconductivity coexist thanks to the formation of pairs with 
finite center-of-mass momentum leading to a spatially oscillating  order parameter. 
The existence of such a phase has remained controversial in dimensions higher than one
in theoretical studies \cite{casalbuoni_nardulli_04,sheehy10,chevy10},
while experiments have  found no  evidence of the FFLO phase in 
three-dimensional systems \cite{partridge06,zwierlein06}.

An important challenge in ultracold fermion experiments, which may have already realized the FFLO state \cite{liao_rittner_10},
is to confirm the existence of FFLO correlations (for recent proposals see, {\it e.g.}, \cite{edge09,roscilde09,lutchyn11,kajala11}). 
A direct measurement of the pair momentum distribution function (MDF) in the partially polarized state 
\cite{feiguin07,casula08,rizzi08} has been suggested to provide such  evidence \cite{yang05}. However, this remains very 
difficult because after turning off all confining potentials, the transverse expansion (in the directions of very tight confinement) 
dominates over the longitudinal one \cite{footnote1}. 
Another interesting possibility is to let the gas expand in the 1D lattice after turning off the longitudinal confining potential, and then measure the density profiles or the MDFs of the independent species and/or pairs after some expansion time. 
Some aspects of such an 
expansion experiment have already been successfully carried out in 1D tubes \cite{kinoshita_wenger_04,kinoshita_wenger_06} as well as in 2D and 3D optical lattices \cite{schneider12}, namely the independent control over lattice and the trapping potential  
and the measurement of the density profiles after the expansion. 
For  1D gases, interaction effects 
during the expansion cannot in general be neglected, leading to fundamentally different behavior of observables before and after the gas has expanded. For example, 
the expansion of the Tonks-Girardeau gas in 1D results in a bosonic gas with a fermionic MDF 
\cite{rigol_muramatsu_05eHCBb,minguzzi_gangardt_05,rigol_muramatsu_05eHCBc}, and initially incoherent (insulating) 
states of bosons \cite{rigol_muramatsu_04eHBCa,rodriguez_manmana_06} and fermions \cite{fabian_rigol_08} can develop quasi-long 
range correlations during the expansion.

The question we are set to address  is the fate of the MDFs of fermions and pairs during an expansion in one dimension, as described by the attractive Hubbard model.  
We use a combination of numerical simulations, based on the time-dependent density matrix renormalization 
group approach ($t$-DMRG) \cite{daley04,white04}, and analytical (Bethe-Ansatz) results. 
We first show that the 
MDFs of majority and minority fermions become stationary after a relatively short expansion time, $t\sim L_0 /J$, where $L_0$ is the 
initial size of the cloud and $J$ is the hopping amplitude. For strong interactions, we  explain this behavior in terms of a quantum distillation process \cite{hm09},
as a consequence of which FFLO correlations are  destroyed during the expansion.
Finally, we  discuss how these stationary MDFs can be theoretically understood within the framework 
of the Bethe-Ansatz. Our results suggest that the final form of the MDFs of minority and majority fermions are related to the 
distributions of Bethe-Ansatz rapidities (a full set of conserved quantities) of this integrable lattice system.

 The Hubbard model (in standard notation \cite{essler}) reads:
\begin{equation}
H_0 = -J\sum_{\ell=1}^{L-1} ( c_{\ell+1,\sigma}^{\dagger}  c_{\ell,\sigma}^{ }
+ \mathrm{H.c.} ) + U \sum_{\ell=1}^{L}
n_{\ell\uparrow} n_{\ell\downarrow} \,. \label{eq:ham}
\end{equation}
As the initial state, we always take the ground state of a trapped system. In the main text, we focus on 
a box trap, {\it i.e.}, particles confined into a region of length $L_0$ while we present results for the expansion
from a harmonic trap in the supplementary material \cite{supp-mat}. 
We study lattices with $L$ sites, $N$ particles, and a global polarization 
of $p=(N_{\uparrow}-N_{\downarrow})/N$, where $N_{\sigma}=\sum_\ell \langle n_{\ell\sigma}\rangle$. All positions are given in units 
of the lattice spacing and momenta in inverse units of the lattice spacing ($\hbar=1$).

The expansion is triggered by suddenly turning off the confining potential, thus allowing particles to expand in the lattice. 
We then follow the time-evolution using the numerically exact {\it t}-DMRG algorithm \cite{daley04,white04}. 
We use a Krylov-space based time-evolution method and enforce discarded weights of $10^{-4}$ or smaller with a time-step of 
$\delta t=0.25/J$. Our main focus is on the time-evolution of the three MDFs: the ones for majority ($\sigma=\,\uparrow$) and 
minority fermions ($\sigma=\,\downarrow$), denoted by $n_{k,\sigma}$ and the pair MDF, $n_{k,{\mathrm{p}}}$. These functions are 
computed from the corresponding one-particle ($\lambda=\uparrow,\downarrow$) or one-pair ($\lambda=p$) density matrices via a Fourier transform
\begin{equation}
n_{k,\lambda} =\frac{1}{L} \sum_{\ell,m} e^{{\rm i}(\ell-m)k} \langle\psi_{\ell,\lambda}^{\dagger} \psi_{m,\lambda}\rangle
\end{equation}
where $\psi_{\ell,\sigma}^{\dagger}=c_{\ell,\sigma}^{\dagger}$,  
$\psi_{\ell,p}^{\dagger}=c^{\dagger}_{\ell,\uparrow}c^{\dagger}_{\ell,\downarrow}$
and $\lambda$ stands for $\uparrow,\downarrow,p$. We normalize the MDFs so that $\sum_k n_{k,\lambda}=N_{\lambda}$ (note that $N_p= \sum_{\ell} \langle  n_{\ell\uparrow}  n_{\ell \downarrow}\rangle$, {\it i.e.}, it  is equal to the 
total double occupancy in the system). 

For the  expansion from a box, we focus on an initial density  fixed to $n=N/L_0=0.8$.
In our {\it t}-DMRG simulations, which were carried out for $N=8$ and $N=16$ ($L_0=10$ and 20, respectively) and various values of $U$, 
we were able to reach times of order $t_{\mathrm{max}}\sim 80/J$ for large $U$ and $t_{\mathrm{max}}\sim 40/J$ for intermediate 
values of $U\sim 4J$. $t_{\mathrm{max}}$ also depends on $p$, with small values of $p$ being more demanding. 

\begin{figure}[t]
\includegraphics[width=0.39\textwidth]{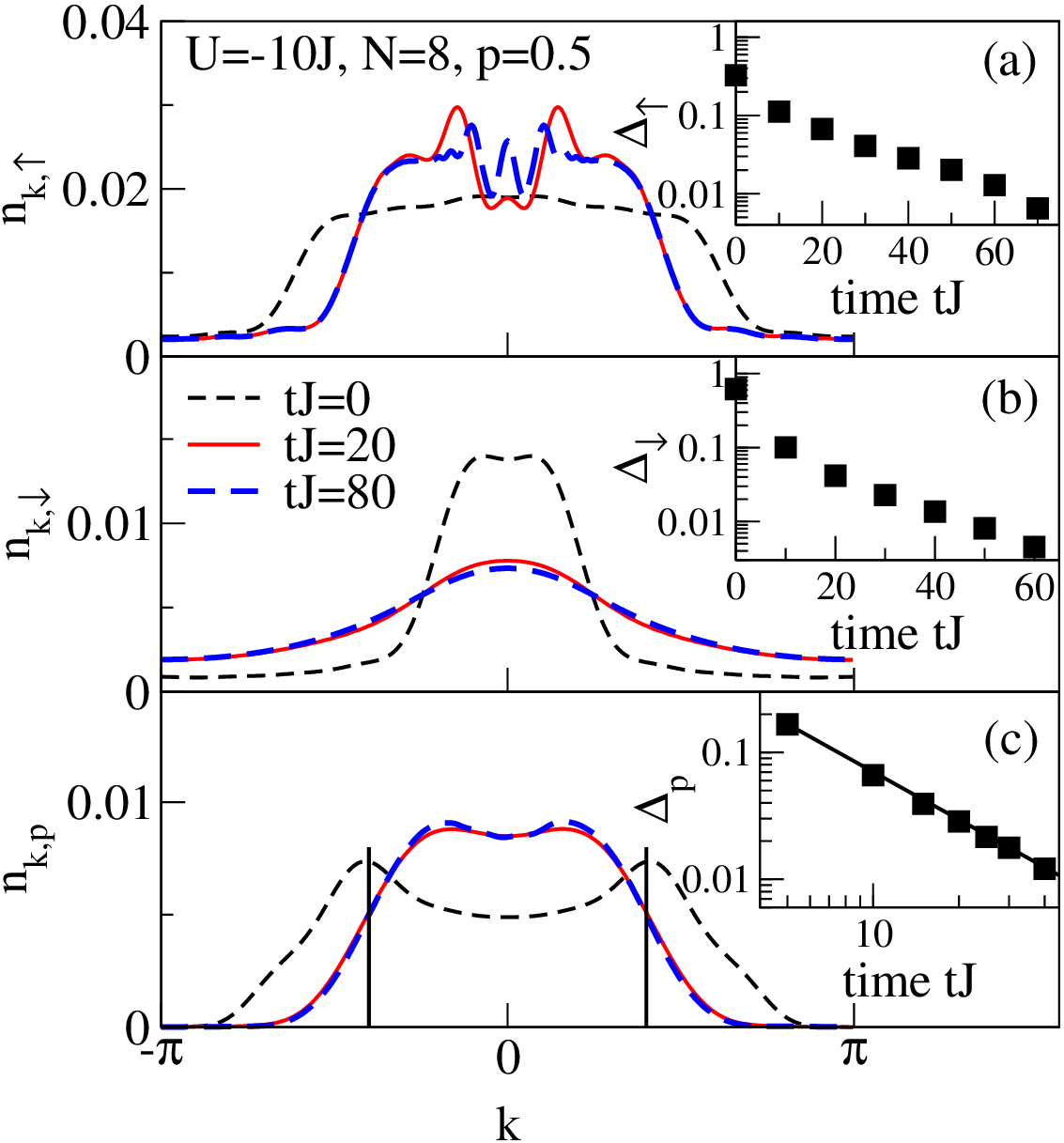}
\caption{(Color online)
MDF for the expansion from a box trap ($U=-10J$, $N=8$, $p=0.5$, $L_0=10$): (a) $n_{k,\uparrow}$, (b)  $n_{k,\downarrow}$,
and (c)  $n_{k,p}$. The insets show the difference $\Delta_{\lambda}$ ($\lambda=\uparrow,\downarrow,p$, see text) between the MDF at a time 
$t$ compared to the one at the largest time reached in the simulation.
The vertical lines in the main panel in (c) mark the position of the FFLO wave-vector $Q=\pm \pi n p$.}
\label{fig:mdf}
\end{figure}

Typical results for the three MDFs of interest are presented in Fig.~\ref{fig:mdf} for $U=-10J$ and $p=0.5$  (corresponding to $N_{\uparrow}=6$ and $N_\downarrow=2$; see the supplementary 
material for more results \cite{supp-mat}). During the time evolution, they are all seen to quickly approach
time-independent forms. In Fig.~\ref{fig:mdf}(a), it is apparent that the MDF of the majority fermions becomes narrower and 
develops small oscillations in the vicinity of $k=0$ as time passes. We find that those oscillations become smaller in amplitude 
and get restricted to smaller values of $k$ after long expansion times, {\it i.e.}, they seem to be a transient feature not  present 
in the asymptotic distributions. The momentum distribution of the minority fermions [Fig.~\ref{fig:mdf}(b)], on the other hand, becomes 
broader during the time evolution.

The time evolution of the MDF of the pairs, depicted in Fig.~\ref{fig:mdf}(c), yields 
information on the fate of FFLO correlations in the expanding cloud. In the FFLO state, $n_{k,p}$ has maxima at 
$Q= \pm(k_{F\uparrow}-k_{F\downarrow})$ \cite{feiguin07}. These are visible in the $t=0$ curve (dashed line), where $\pm Q$
are marked by vertical lines. As the comparison of $n_{k,p}(t>0)$ with the initial $n_{k,p}(t=0)$ shows, the peaks at $\pm Q$  
rapidly disappear, and $n_{k,p}(t)$ becomes narrower. In addition, new and shallower peaks form at $k<Q$. Since we do not find 
those peaks at the same values of $k$ for other values of $N$ when $N/L_0$ and $p$ are the same, and we do not find them 
for all values of $U$, $N/L_0$, and $p$ studied, they appear to be related to finite-size effects. Hence, the double peak 
structure in $n_{k,p}(t=0)$, which makes evident the presence of FFLO correlations in the initial state, is found to 
disappear during the expansion.
Even though  the FFLO correlations are lost during the expansion, the integral over the pair MDF, which equals the total
double occupancy, does not vanish.
This implies  that not all interaction energy is converted into kinetic energy and that some fraction of the original pairs remains by the time
the MDFs have become stationary, which in experiments could be probed by measuring the double occupancy. 

In order to quantify how the three MDFs above approach stationary forms, in the insets in Fig.~\ref{fig:mdf}, we plot 
$\Delta_{\lambda}(t) = \sum_k |n_{k,\lambda}(t) - n_{k,\lambda}(t_{\mathrm{max}})|/\sum_k n_{k,\lambda}(t_{\mathrm{max}})$ vs $t$. 
These results make apparent that the approach is close to exponential for $n_{k,\uparrow}$ and $n_{k,\downarrow}$ 
[insets in Fig.~\ref{fig:mdf}(a) and \ref{fig:mdf}(b)], while it is power law for $n_{k,p}$ [inset in Fig.~\ref{fig:mdf}(c)] \cite{power}. 
Remarkably, for the parameters of Fig.~\ref{fig:mdf}, already at $tJ\sim10$, all $\Delta_{\lambda}$ are $\lesssim 10\%$. This means that the 
stationary MDFs obtained in this work should be achievable in current optical lattice setups \cite{schneider12}. A comparison 
between expansions from different box sizes suggests that the emerging time scale in the observables with exponential relaxation 
is proportional to $L_0$. The origin of that time scale will be  discussed below. 

While we are focusing the discussion on the case of the expansion from a box trap, we stress that  the results for the MDFs in 
an expansion from a harmonic trap are quite similar  
(for an example, see the supplementary material \cite{supp-mat}). Namely, we observe 
a comparably fast convergence of the MDFs to a stationary form 
and the disappearance of the peaks at $\pm Q$  in $n_{k,p}$. The latter 
indicates the disappearance of FFLO correlations.

\begin{figure}[!t]
\includegraphics[width=0.48\textwidth]{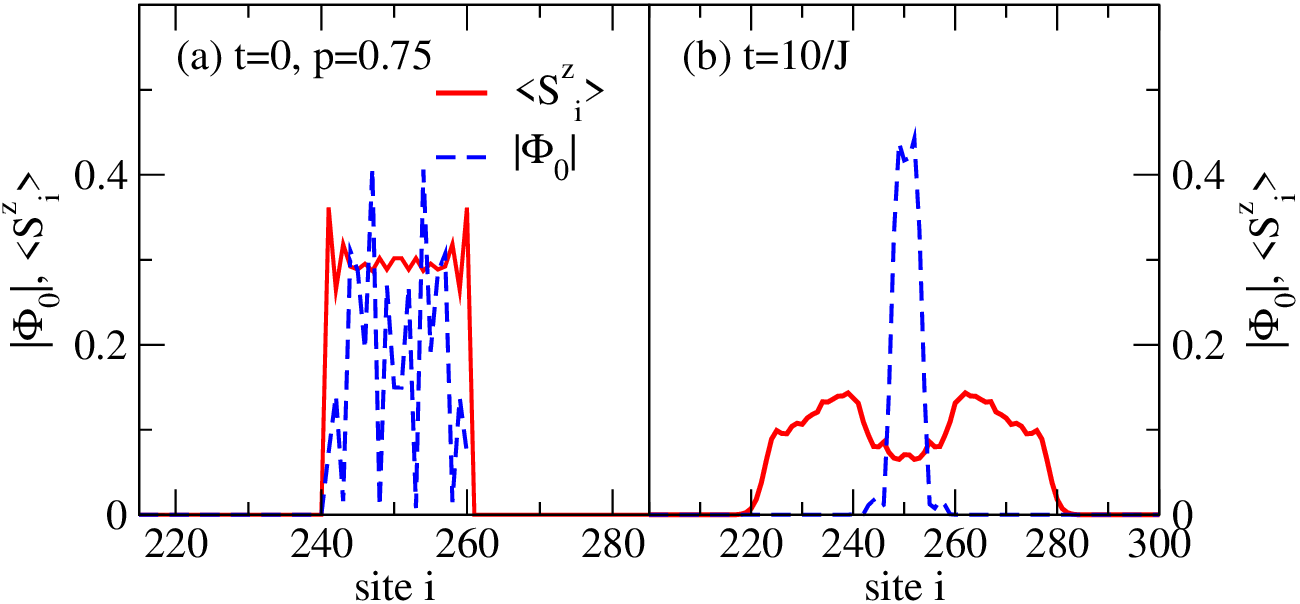}
\caption{(Color online) Natural orbital $|\Phi_0|$ corresponding to the largest eigenvalue of the pair-pair correlator $P(\ell,j)$
(dashed lines) and spin density $\langle S_i^z\rangle $ (solid lines).
(a) $t=0$, (b) $tJ=10$. These results are for $U=-10J$,  $L_0=20$,  $N=16$ and $p=0.75$, corresponding to $N_{\uparrow}=14$ and $N_{\downarrow}=2$.}
\label{fig:opdm}
\end{figure}

To understand how the FFLO state breaks down as the gas expands, we calculate the eigenvector $\Phi_{0}$ of the pair-pair 
correlator $P(\ell,m)=\langle \psi_{\ell,p}^{\dagger} \psi_{m,p}\rangle$ that corresponds to the largest eigenvalue.
$|\Phi_{0}|$, shown in Fig.~\ref{fig:opdm}(a), unveils the spatial structure of the quasi-condensate in the initial state: 
it has an oscillatory behavior with nodes (see also Ref.~\cite{feiguin07}). In these nodes, the spin density has its maxima to 
accommodate the majority fermions (Fig.~\ref{fig:opdm}(a), see also \cite{supp-mat}), indicative of the spin-density wave character with a modulation 
of $(2Q)^{-1}$ in the FFLO state. During the expansion, the nodes in $|\Phi_{0}|$ disappear while $|\Phi_{0}|$ 
develops a maximum at $L/2$, exceeding its initial value [see Fig.~\ref{fig:opdm}(b)]. The latter is a consequence 
of a quantum distillation mechanism, described in Ref.~\cite{hm09} for $U>0$, which allows the unpaired fermions to move away from
the center of the system ({\it i.e.}, they escape from the nodes of $|\Phi_{0}(t=0)|$). Loosely speaking, during first-order processes  
unpaired fermions exchange their positions with the pairs (a minority fermion hops towards the center of the trap), allowing the former 
to expand while the pairs move towards the center of the trap. This occurs over a time scale proportional to $L_0$ and inversely 
proportional to $J$, which explains the time scale observed in the exponential approach of the majority and minority fermions 
to their stationary values. Once the unpaired fermions have spatially separated themselves from the pairs, they form a non-interacting gas 
whose MDF is stationary. On much longer time scales (assuming $U>4J$), we expect the pairs to slowly expand as well. This transient dynamics of the pairs may 
be the reason for the power-law, as opposed to exponential, relaxation observed for $n_{k,p}(t)$ in Fig.~\ref{fig:mdf}(c).

In a recent work \cite{lu12}, extrema in the spin-density of the expanding gas were observed in numerical calculations
using various approaches. By comparing with the time-dependence of the order parameter within a time-dependent Bogoliubov-deGennes
approach, it was argued that they are related to FFLO correlations. Our results show that, in a lattice system,
the nodal structure of the FFLO state is ultimately lost as the system expands. Note, however, that in Ref.~\cite{lu12} the main focus was on rather small polarizations $p$  \cite{orso07,hu07,hm10} leading to a wide
partially polarized core before the expansion. We therefore expect the quantum distillation mechanism to take much longer to depolarize the core than what has so far been
reached in numerical simulations \cite{lu12}, leaving this case as an open question.

We are now in a position to explain the anticorrelated behavior of $n_{k,\uparrow}$ and $n_{k,\downarrow}$ mentioned in the discussion of
Fig.~\ref{fig:mdf}.
For large values of $U$, $N_p$ is essentially equal to $N_{\downarrow}$ and is
approximately unchanged during the expansion, rendering the interaction energy
almost time independent.
This implies that also the kinetic energy $E_\textrm{kin}=-2J\sum_k \cos k (n_{k,\uparrow}+n_{k,\downarrow})$ is
approximately conserved, which is only possible if the two MDFs behave in the opposite way during the expansion.
The broadening of the minority MDF $n_{k,\downarrow}$ with respect to the initial state is a direct consequence of the spatial separation of excess 
fermions from the pairs,
leaving the latter confined in the center of the cloud. Since in the center the local polarization decreases, the stationary form of  $n_{k,\downarrow}$ is well approximated by the equilibrium one for equal populations $N_\uparrow=N_\downarrow$
instead of  $N_\uparrow >N_\downarrow$ \cite{supp-mat}.

The fact that the MDFs become stationary after the expansion from a box or a harmonic trap is in itself not surprising, as 
in the limit of long expansion times, the cloud becomes very dilute with, for the attractive case, the typical inter-particle 
distance being much larger than the bound-state size. Hence, one may assume that pairs and unpaired particles are essentially 
noninteracting. The MDF in such an asymptotic limit should be determined by the initial conditions right after the quench. 
For instance, for generic models, the total energy (which is conserved during the expansion) plays a fundamental role
in determining the expansion dynamics (see Ref.~\cite{langer12} for a related work for $U>0$). For an integrable model, such 
as the (attractive) Hubbard model of Eq.~\eqref{eq:ham}, all integrals of motion are in principle known from the Bethe Ansatz 
and are conserved during the expansion \cite{essler}. We  argue below how to interpret the shape of certain stationary MDFs 
in terms of such integrals of motion. This is closely related to the previously studied fermionization of the MDF of an expanding 
gas of hard-core bosons \cite{rigol_muramatsu_05eHCBb,minguzzi_gangardt_05,rigol_muramatsu_05eHCBc}. 

For the model studied here, we first note that the formation of a distinct minimum in the difference distribution 
$\delta n_k=n_{k,\uparrow}-n_{k,\downarrow}$ [see Figs.~\ref{fig:comp_mdf}(a) and (b)] is reminiscent of the corresponding distribution of real-valued charge rapidities 
(for intermediate $U$) in  the ground state in a box.
From the point of view of the rapidity distributions, they need to be determined right after turning off the trap 
and the subsequent expansion does 
not play any role; it is the MDFs which will evolve and asymptotically approach the former as the expansion proceeds 
\cite{sutherland_98}. We can calculate the pre-quench values of the rapidities 
by numerically solving the Bethe-Ansatz equations 
for a system of size $L_0$ and open boundary conditions \cite{sklyanin88,zhou96,guan00}. For the ground state of the attractive Hubbard model, 
there are two types of rapidities present: real- and complex-valued charge rapidities ($\kappa_{\nu}$ and $\kappa_{\sigma}$) which correspond to unpaired fermions and 
pairs, respectively ($\nu=1,\dots, N_{\uparrow}-N_{\downarrow}$, $\sigma=1,\dots,2N_{\downarrow}$, with $\kappa_{\sigma}$ and $\kappa^*_{\sigma}$ appearing pairwise). 

To calculate the effect of the  quench of the trapping potential exactly is in principle possible but complicated in practice 
\cite{mossel10}, so we will resort to  some simplifications. To start, we assume that the number of pairs is conserved 
during the quench, and thus no pure-spin excitations are produced. Further, we use the observation that the overlap between the 
pre-quench eigenstate and the post-quench state has a maximum amplitude for components of the latter with the same set of 
rapidities \cite{mossel10}. We then identify, asymptotically, the distribution of real-valued charge rapidities with that of 
unpaired fermions ($\delta n_k$), 
and of the real part of complex-valued (string) charge rapidities with that of minority fermions ($n_{k,\downarrow}$) since they 
remain paired. Finally, we model the quench by convolving the pre-quench distributions $\rho_{1}=(1/2)\sum_{\nu} \delta{(k\pm \kappa_{\nu})}$ and $\rho_2=(1/2)\sum_{\sigma} \delta{(k\pm \mathrm{Re} \kappa_{\sigma})}$ with the (periodized) kernels: 
(i) $L_0\,\mathrm{sinc}^2( k L_0/2)$ for the former and (ii) a simple Lorentzian for the latter. The first choice 
is inspired by the exact result for the release of a single particle from a box, while the second choice is done for simplicity given 
that the results are relatively featureless in comparison. Illustrative results are shown in Fig.~\ref{fig:comp_mdf} and the agreement 
is very good, specially away from the Brillouin-zone center. Note that there are no fitting parameters in the case of $\delta n_k$ and 
a single fitting parameter, the width of the Lorentzian, in the case of $n_{k,\downarrow}$.

\begin{figure}[t]
\includegraphics[width=0.48\textwidth]{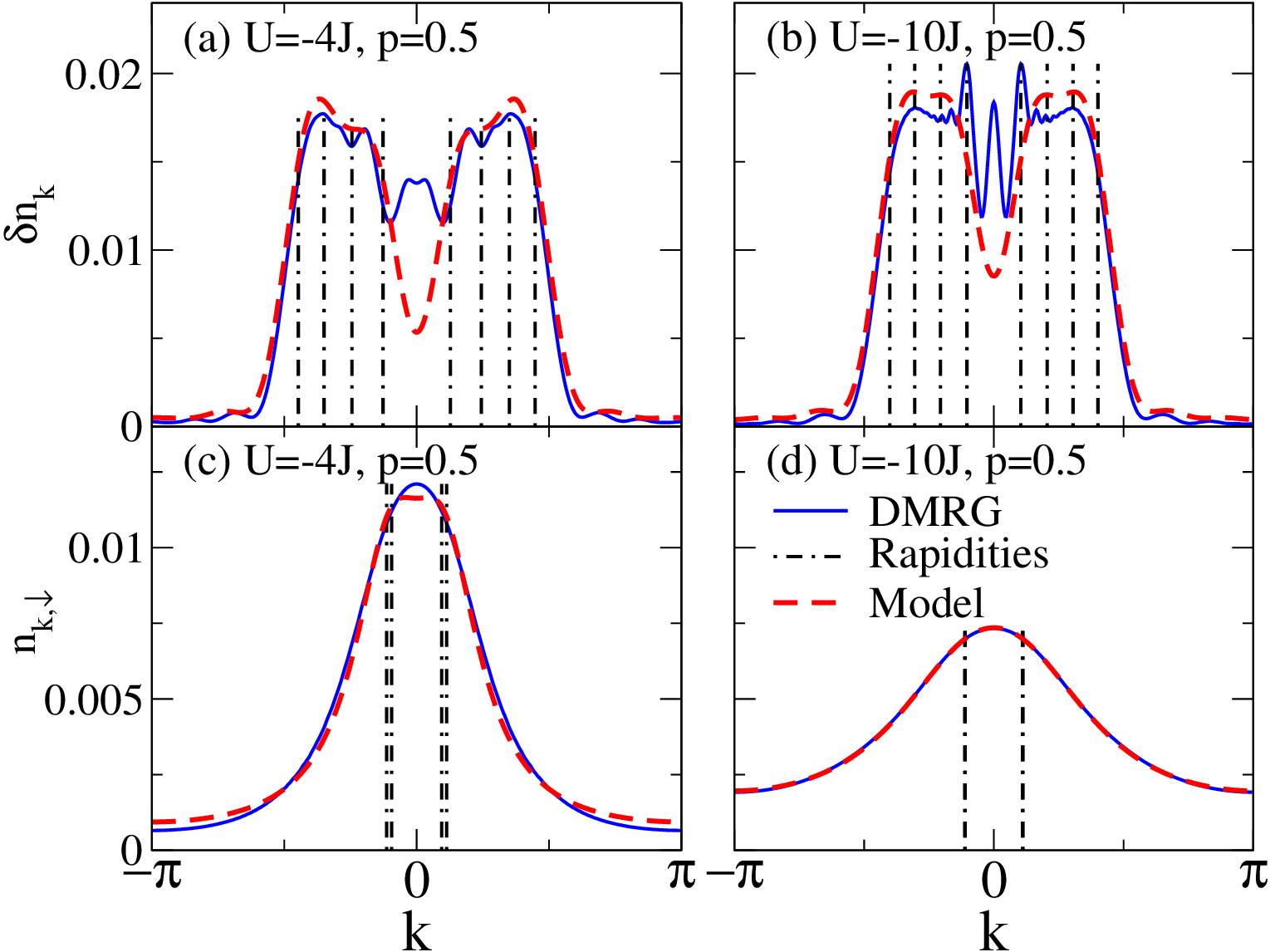}
\caption{(Color online) Comparison of the stationary MDFs $\delta n_k=n_{k,\uparrow}-n_{k,\downarrow}$  [(a),(b)] and $n_{k,\downarrow}$ 
[(c),(d)] for the expansion from a box with $N=8$, $p=0.5$ (corresponding to $N_{\uparrow}=6$, $N_{\downarrow}=2$) [(a),(c): $U=-4J$, (b),(d): $U=-10J$] to the form expected from 
the rapidities known from the Bethe-Ansatz: $t$-DMRG (solid lines), models  discussed in the text (dashed lines). 
The vertical lines mark the positions of the rapidities.}
\label{fig:comp_mdf}
\end{figure}

In conclusion, we demonstrated that the initial FFLO state is destroyed during
the expansion of an attractively interacting partially polarized 1D Fermi gas, and that direct signatures of the FFLO phase in the initial pair
MDF are washed out as a consequence of interactions. 
Nevertheless, the sudden expansion is an interesting non-equilibrium experiment that
through the asymptotic form of the MDFs 
yields information on the initial state. 
Our analysis suggests that the shape of the MDFs can be related to the
distribution of rapidities, which constitute a full set of integrals of motion
for this integrable quantum model and fully determine the initial state.
Since we showed that the MDFs 
of majority and minority fermions as well as the one of pairs
rapidly take a stationary form, this should be accessible on typical experimental time-scales.

\begin{acknowledgments}
We thank X. Guan and R. Hulet for their comments on the manuscript. We thank the KITP at UCSB, where this work was initiated, for its hospitality and NSF for support under grant No.~PHY05-51164 
(CJB, FHM, and MR). We also acknowledge support from the DARPA OLE program through 
ARO W911NF-07-1-0464 (CJB), the Deutsche Forschungsgemeinschaft through FOR 801 (FHM and SL) and FOR 912 (SL), 
and the Office of Naval Research (MR). 
I.P.M. acknowledges support from the Australian Research Council
Centre of Excellence for Engineered Quantum Systems and the 
Discovery Projects funding scheme (project number DP1092513).\end{acknowledgments}

\pagebreak 

\title{Electronic physics auxiliary publication service for: Long-time behavior of the  momentum distribution during the sudden expansion of \\
a spin-imbalanced Fermi gas in one dimension
}

\author{C. J. Bolech}
\affiliation{Department of Physics, University of Cincinnati, Cincinnati, OH 45221, USA}
\affiliation{Kavli Institute for Theoretical Physics, Kohn Hall, University of California, Santa Barbara, CA 93106}
\author{F. Heidrich-Meisner}
\affiliation{Kavli Institute for Theoretical Physics, Kohn Hall, University of California, Santa Barbara, CA 93106}
\affiliation{Department of Physics and Arnold Sommerfeld Center for Theoretical Physics, 
Ludwig-Maximilians-Universit\"at M\"unchen, D-80333 M\"unchen, Germany}
\author{S. Langer}
\affiliation{Department of Physics and Arnold Sommerfeld Center for Theoretical Physics, 
Ludwig-Maximilians-Universit\"at M\"unchen, D-80333 M\"unchen, Germany}
\author{I. P. McCulloch}
\affiliation{School of Physical Sciences, The University of Queensland, Brisbane, QLD 4072, Australia}
\author{G. Orso}
\affiliation{Laboratoire Mat\'eriaux et Ph\'enom\'enes Quantiques, Universit\'e Paris Diderot-Paris 7 and CNRS, 
UMR 7162, 75205 Paris Cedex 13, France}
\author{M. Rigol}
\affiliation{Department of Physics, Georgetown University, Washington, DC 20057, USA}
\affiliation{Physics Department, The Pennsylvania State University, 104 Davey Laboratory,
University Park, Pennsylvania 16802, USA}

\maketitle
\date{July 30, 2012}

\section*{Additional results for the MDFs of a spin-imbalanced Fermi gas with attractive interactions}
\begin{figure}[bh]
      \center
    \includegraphics[width=0.8\linewidth]{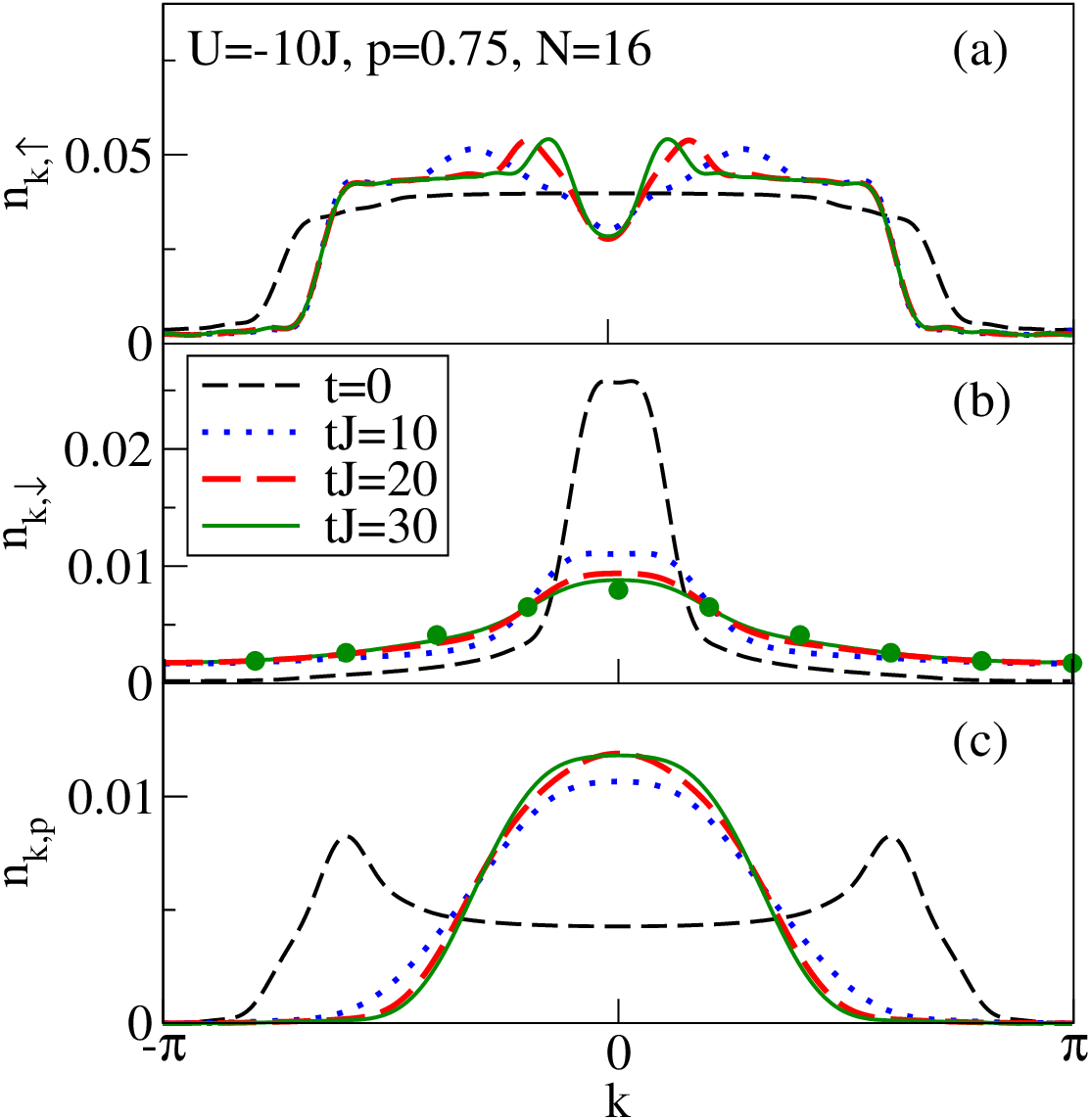}    
     \caption{
    MDFs for (a) spin up, (b) spin down, and (c) pairs for the expansion from a box trap
    with $U=-10J$, $N=16$, $p=0.75$, $L_0=20$, plotted at times $tJ=0,10,20,30$. The circles
    in the central panel represent the MDF of an unpolarized gas with 
    $N_\uparrow=N_\downarrow=2$.
	}\label{fig:coarse}
\end{figure}

We here provide additional $t$-DMRG results for the MDFs $n_{k,\lambda}$ 
of a spin-imbalance Fermi gas with attractive interactions, expanding from a box trap. 
Figure~\ref{fig:coarse}
contains data for $N=16$ (with $U=-10J$, $p=0.75$, $L_0=20$). In this case we were
able to reach maximum times of $t_{\mathrm{max}} \sim 30/J$. Nevertheless, a fast approach
towards a stationary form is obvious from this figure, corroborating the 
conclusions of the main text (see the discussion of Fig.~1 of the main text). The same
applies to the qualitative trends: $n_{k,\uparrow}$ shrinks while $n_{k,\downarrow}$
broadens.

\begin{figure}[t]
      \center
    \includegraphics[width=0.8\linewidth]{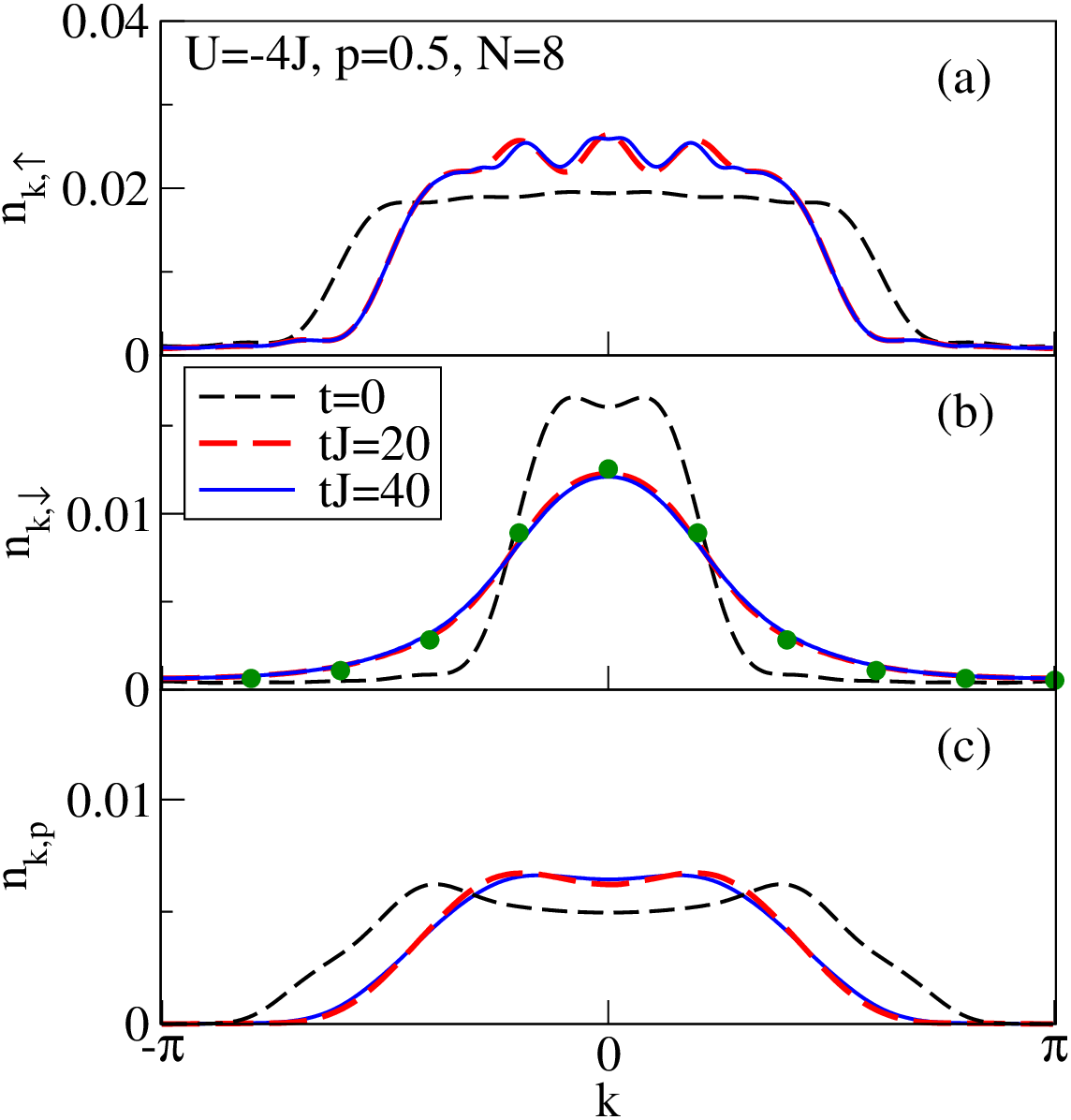}
     \caption{
    MDFs for (a) spin up, (b) spin down, and (c) pairs for the expansion from a box trap
    with $U=-4J$, $N=8$, $p=0.5$, $L_0=10$, plotted at times $tJ=0,20,40$. The circles
    in the central panel represent the MDF of an unpolarized gas with 
    $N_\uparrow=N_\downarrow=2$.
    }\label{fig:U-4}
\end{figure}

In Fig.~\ref{fig:U-4}, we display results for $U=-4J$ and $N=8$ with a polarization of 
$p=0.5$. In this case, the convergence to a stationary form is evident in all three 
MDFs. Note that, in contrast to the case of $U=-10J$ discussed in the main text,
$n_{k,\uparrow}$ does not exhibit any transient fluctuations at small $k$.

\section*{Discussion of the qualitative behavior of the MDFs of the spin-imbalanced
Fermi gas with attractive interactions}
Comparing  Figs.~1(a)-(b) of the main text, as well as Fig.~\ref{fig:U-4}(a) and (b) shown here in the supplementary material, we see that the momentum distribution $n_{k,\uparrow}$ of the majority component 
shrinks during the expansion, whereas the  distribution $n_{k,\downarrow}$ of the minority component 
broadens significantly. This is the result of collisions between up and down fermions which take place in the inner part  of the system, where both spin components are present, and transfer momenta between them.  
In the long time limit and for $|U|>J$, the pairs phase separate from unbound fermions, such that the cloud develops a two-shell structure with a fully paired core  and fully polarized wings containing the excess $N_\uparrow-N_\downarrow$ fermions. 
This suggests that the asymptotic momentum distribution of the minority component should be approximated by its  ground-state value before the expansion calculated \textsl{in the absence} of excess fermions, that is for $N_\uparrow=N_\downarrow$. The results for the MDF $n_{k,\downarrow}$ that we obtain using this assumption  are plotted in Figs.~\ref{fig:coarse} and \ref{fig:U-4}  with circles.
Indeed, we  see a rather good agreement with the stationary form of the MDF $n_{k,\downarrow}$, where the latter was calculated with $t$-DMRG. In particular, in the limit of 
 large initial polarization $p\rightarrow 1$, the number of pairs is very small. In this low density (or equivalently, strong-coupling) regime the ground state momentum distribution becomes equal to the
square of the Fourier transform of the molecular wave-function for the relative motion:
\begin{equation}
n_{k\downarrow}=n_{\downarrow}
 \frac{|U|^3}{\sqrt{U^2+16J^2}}
 \frac{1}{(-4J \cos k+\sqrt{U^2+16J^2})^2}.
\end{equation}
The corresponding shrinking of the majority momentum distribution  during the expansion can then be understood from conservation of total energy. Indeed, for $|U|\gg J$ the number of double occupancies remains close to 
$N_p=N_\downarrow$ during the expansion, implying that the interaction energy in the Hubbard model is essentially time independent. As a consequence, the kinetic energy $E_\textrm{kin}=-2J \sum_k \cos k (n_{k,\uparrow}+n_{k,\downarrow})$ is also  conserved, implying that the distribution $n_{k,\uparrow}$ must shrink to compensate the broadening of $n_{k,\downarrow}$.

\begin{figure}[t]
    \centerline{\includegraphics[width=0.8\linewidth]{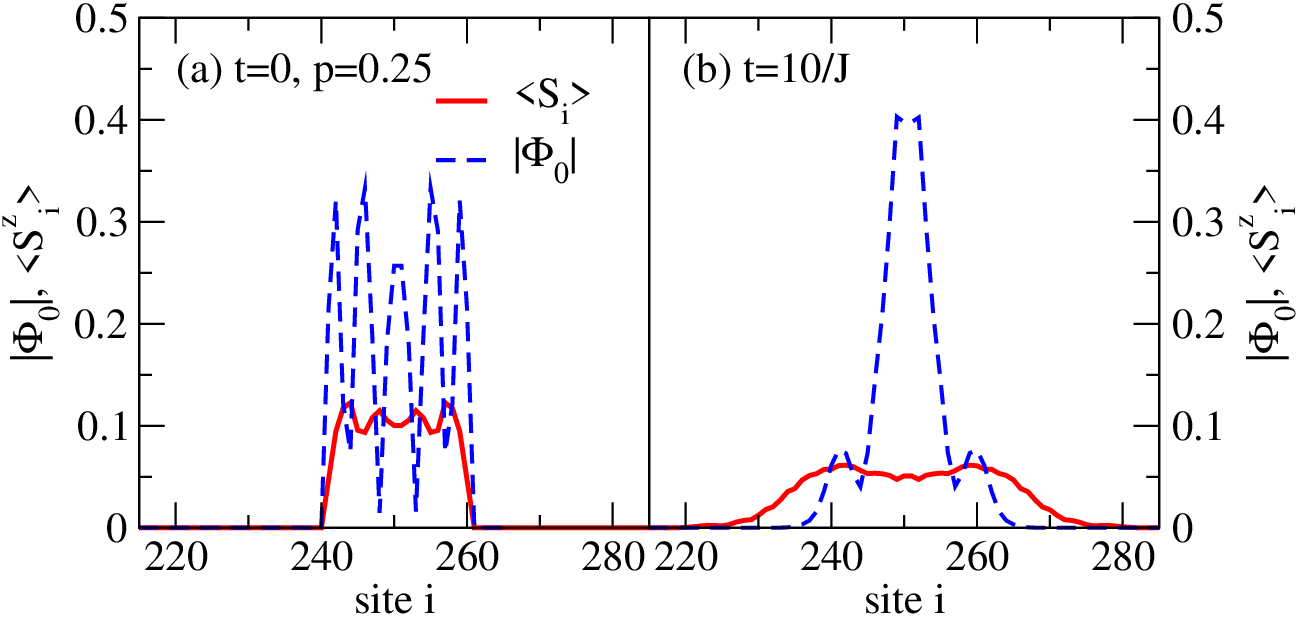}}
    \centerline{\includegraphics[width=0.80\linewidth]{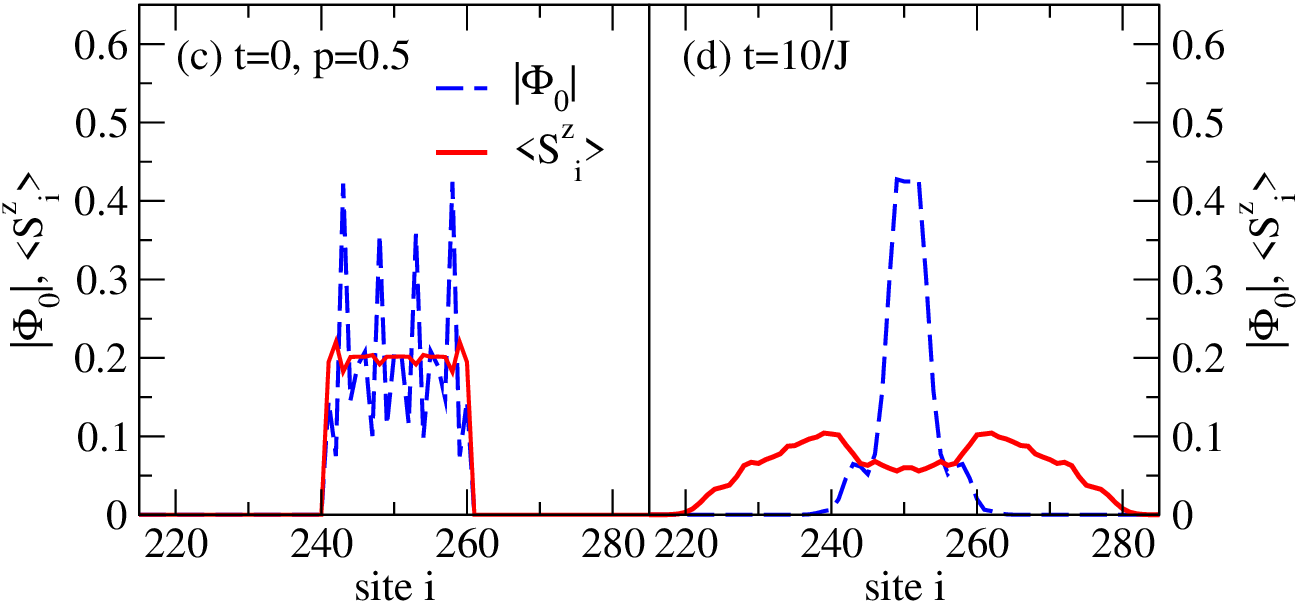}}
     \caption{ Natural orbital $|\Phi_0|$ corresponding to the largest eigenvalue of the pair-pair correlator $P(\ell,j)$
(dashed lines) and spin density $\langle S_i^z\rangle $ (solid lines).
(a),(c) $t=0$, (b),(d) $tJ=10$. These results are for $U=-10J$,  $L_0=20$,  $N=16$ and $p=0.25$ [panels (a) and (b)],
and p=0.5 [panels (c) and (d)].}
        \label{fig:opdm_s}
\end{figure}

\section*{Additional results for the time-evolution of the quasicondensate}
In Fig.~2 of the main text, we show the spin-density $\langle S_i^z\rangle$ and the eigenvector $|\Phi_0|$ of the 
pair-pair correlator in the initial state and after an expansion time of $tJ=10$ for $U=-10J$ and $p=0.75$ 
(corresponding to $N_{\uparrow}=14$ and $N_{\downarrow}=2$ with $L_0=20$). Here we present additional results
for $p=0.25$ ($N_{\uparrow}=10$ and $N_{\downarrow}=6$) and ($N_{\uparrow}=12$ and $N_{\downarrow}=4$) 
in Fig.~\ref{fig:opdm_s}. The nodal structure of quasicondensate is best seen in the case of $p=0.25$ [Fig.~\ref{fig:opdm_s}(a)]  
with the spin-density taking maxima in the nodes of $|\Phi_0|$.
During the expansion, we observe the same spatial demixing of excess fermions from the pairs that is discussed in the 
main text for the case of $p=0.75$.
Note that the demixing occurs over short expansion times during which the cloud has expanded by 
about a factor of two only.

\begin{figure}[b]
    \centerline{\includegraphics[width=0.8\linewidth]{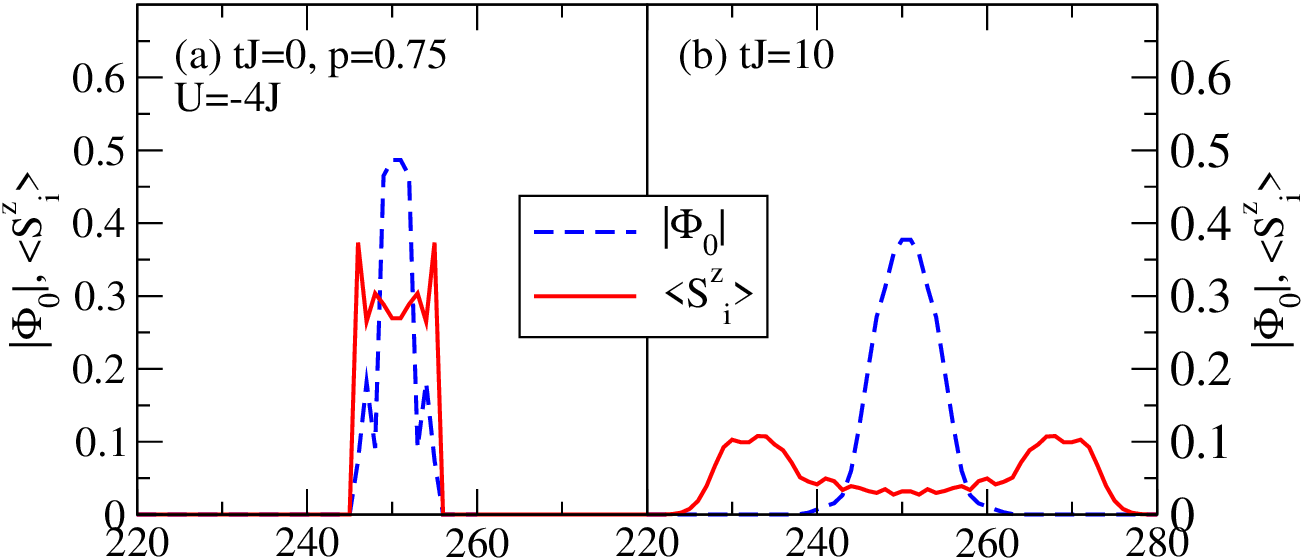}}
     \caption{ Natural orbital $|\Phi_0|$ corresponding to the largest eigenvalue of the pair-pair correlator $P(\ell,j)$
(dashed lines) and spin density $\langle S_i^z\rangle $ (solid lines).
(a) $t=0$, (b) $tJ=10$. These results are for $U=-4J$,  $L_0=10$,  $N=8$ and $p=0.75$.
}
        \label{fig:opdm4}
\end{figure}

In Fig.~\ref{fig:opdm4} we demonstrate that the demixing is not limited to the strongly interacting regime. We present
results for $\langle S_i^z\rangle$ and  $|\Phi_0|$ for $U=-4J$ at $p=0.75$ with $N=8$. Obviously, at time $tJ=10$,
we observe that the spatial structure of the FFLO-quasicondensate is lost and that excess fermions and pairs
are separated from each other. 
The main reason for the separation of excess fermions from pairs is, as discussed in the main text, the difference in the 
bare velocities of pairs and excess fermions. In the lattice this difference is large at $U\gg J$ yet the quantum
distillation mechanism still works even at intermediate values of $|U|\sim 4J$ as shown here. The only noticeable 
difference between large $U$ and intermediate $U$ is that the density of the pairs in the core of the cloud
does not increase as the system expands at $U=-4J$.  
Based on our results for $U=-4J$, we, therefore, speculate that in the contiuuum, where the bare velocities of excess fermions versus pairs 
differ by a factor of two only,  
the demixing of excess fermions and pairs should also occur during the expansion. However, it is numerically
very demanding to simulate this limit using time-dependent DMRG since very low densities and therefore, long
expansion times would be necessary.

\section*{MDFs for the sudden expansion from a harmonic trap}

\begin{figure}[!t]
\includegraphics[width=0.8\linewidth]{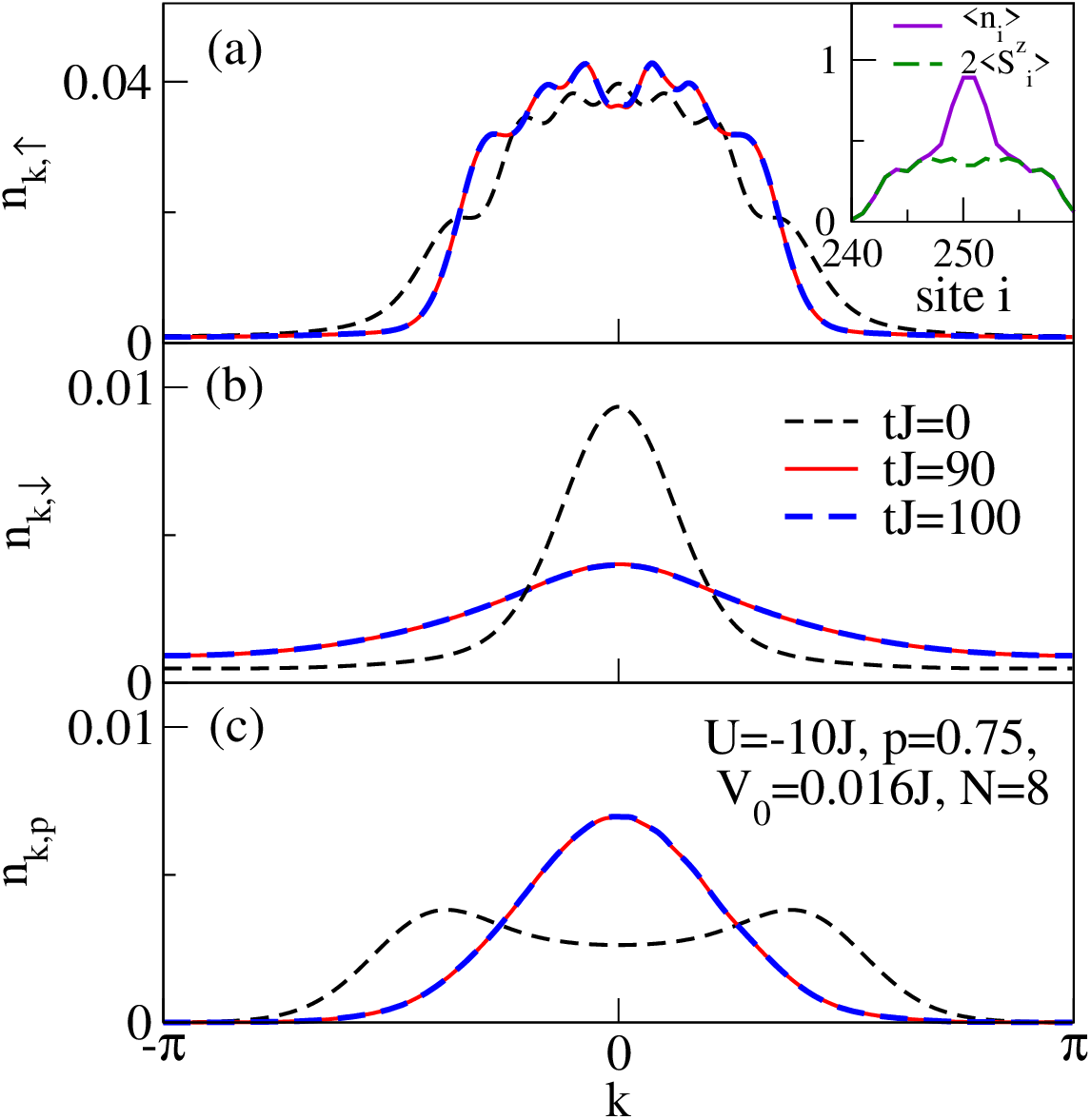}
\caption{(Color online) MDFs for the expansion from a harmonic trap: (a)  $n_{k,\uparrow}$, (b)  $n_{k,\downarrow}$ and (c) $n_{k,p}$.These results were obtained for $N=8$, $U=-10J$, $p=0.75$, $V=0.016J$, and at times $tJ=0,90,100$. Inset in (a): initial density
$\langle n_{i}\rangle$ (solid line) and spin-density profile $2\langle S^z_{i}\rangle$ (dashed line).}
\label{fig:mdf_trap}
\end{figure}

In relation with experiments, it is also important to incorporate the harmonic confinement, {\it i.e.}, $H_{\mathrm{trap}} =V_0 \sum_{\ell=1}^L n_{\ell}\, (\ell-L/2)^2$. To that end, we have prepared a
spin-imbalanced system with $U=-10J$ in a harmonic trap with $V_0>0$ for $t<0$, and then quenched the trapping potential to
$V_0=0$ at $t=0$. For the parameters of Fig.~\ref{fig:mdf_trap},  the partially polarized phase that sits in the core
is surrounded by fully polarized wings (see the inset in Fig.~\ref{fig:mdf_trap}). During the expansion, one can see that the
behavior of the MDFs is very similar to the one starting from a box in Fig.~1 of the main text. All MDFs become stationary shortly
after the release from the trap. The stationary $n_{k,\uparrow}$ is narrower while $n_{k,\downarrow}$ is broader than their corresponding initial distributions, and the double peak structure in $n_{k,p}$, which is due to the FFLO correlations in the initial state,  disappears.

\section*{Time-evolution of the MDFs of a two-component Fermi gas with repulsive interactions}
 
We have also studied the time-evolution of other 1D models during the expansion, including most notably the 
repulsive Hubbard model with $p=0$ (compare \cite{fabian_rigol_08s}).

\begin{figure}[t]
      \center
    \includegraphics[width=0.8\linewidth]{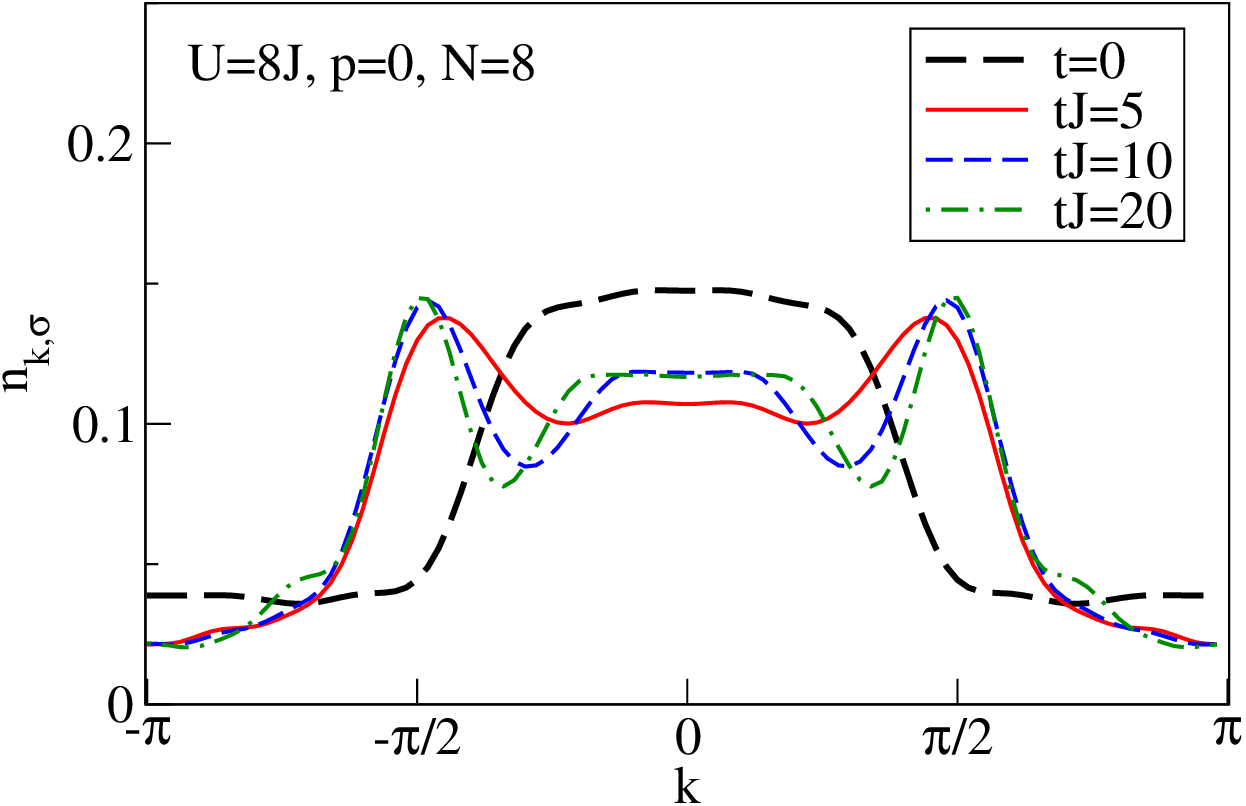}
     \caption{
    MDF $n_{k,\sigma}=n_{k,\uparrow}=n_{k,\downarrow}$  for the expansion from a box trap
    with $U=8J$, $N=8$, $p=0$, plotted at times $tJ=0,5,10,20$. We observe that in the case of repulsive interactions,
    much shorter times can be reached than for $U<0$. On the accessible time scales, the MDF still changes, yet at its edge,
    the curves for $tJ=10$ and $tJ=20$ lie on top of each other.}
        \label{fig:U8}
\end{figure}

 The $U>0$ case turns out to be a numerically much harder problem for $t$-DMRG, as entanglement grows much faster (see the review \cite{schollwoeck11} 
for how entanglement growth limits $t$-DMRG). Therefore, we resorted to exploiting non-Abelian symmetries as well, 
restricting the analysis to $p=0$, allowing us to reach $t\sim 25/J$ for $U=8J$ (see Fig.~\ref{fig:U8}).
In the case of $U=8J$, there are no pairs, and hence over the full extent of the expanding cloud, majority and minority fermions can still 
interact, whereas in the case of $U<0$ and $p>0$, fast majority fermions escape \cite{kajala11s} and the pairs and majority fermions mostly 
decouple due to the quantum distillation mechanism that is described in the main text.
 This is likely the reason why in the  repulsive gas with $U>0$ and $p=0$, there  is a stronger entanglement 
during the expansion. On the accessible time scales, the MDF $n_{k,\sigma}$ of the repulsive gas still undergoes changes, yet in the edge 
of the MDF, the curves at the longest times coincide (see also \cite{fabian_rigol_08s}). The case of $U>0$ thus sets an example where the quantum simulation with ultra-cold 
atomic gases could help us to go to longer times than what is currently possible with numerical methods to clarify the asymptotic behavior 
of the MDF (compare the relaxation dynamics problem studied in Ref.~\cite{trotzky12}).
Note that for the expansion of a repulsive gas with initial densities $\langle n_i\rangle \leq 1$, the double-occupancy {\it decreases} \cite{fabian_rigol_08s}, in contrast
to the attractive case, discussed in the main text. In the attractive case, the survival of a certain fraction of the initial double occupancy
is expected due to the presence of pairs.


\begin{thebibliography}{51}
\bibitem[{foo()}]{footnote1}
 \bibinfo{pages}{This can be understood in terms of Heisenberg uncertainty:
  a tighter confinement implies a larger population of higher
  canonically--conjugate momentum modes, which in turn contribute to a faster
  expansion rate.}
\bibitem[{pow()}]{power}
\bibinfo{pages}{For the exponent of the power--law decay of
  $\Delta_{p}\propto (tJ)^{-\alpha}$, we find $\alpha \approx 1.25$.}
\end{thebibliography}
\end{document}